\documentclass[sigconf,10pt]{acmart}
\usepackage{wrapfig}
\usepackage{comment}
\usepackage{colortbl}

\usepackage{hyperref}

\usepackage{amsthm} 

\usepackage{color,soul}
\definecolor{lightblue}{rgb}{.4,.8,1}
\sethlcolor{lightblue}
\usepackage{graphics}
\usepackage{hyperref}
\usepackage{lipsum}
\usepackage{tikz}
\usepackage{enumitem}	
\usepackage{listings} 
\usepackage{booktabs}	
\usepackage{lipsum}

\usepackage{capt-of}	
\usepackage{diagbox}
\usepackage{multirow}
\usepackage{todonotes}  
\definecolor{refkey}{rgb}{249,158,26}
\definecolor{labelkey}{rgb}{0,1,0}
\definecolor{airforceblue}{rgb}{0.36, 0.54, 0.66}
\definecolor{applegreen}{rgb}{0.55, 0.71, 0.0}

\usepackage{mathrsfs}	
\usepackage{amsfonts}

\usepackage[export]{adjustbox} 

\usepackage[plain]{algorithm2e}
\usepackage{boldline}					

\usepackage{multirow}

\definecolor{frenzyorange}{RGB}{249, 158, 26}
\newcommand*\circled[1]{\tikz[baseline=(char.base)]{
			\node[shape=circle,draw,text=black,inner sep=1pt] (char) {#1};}}
		
\renewcommand{\paragraph}[1]{\vskip 3pt\noindent\textbf{#1 }}	 

  {\begin{list}{$\bullet$}%
     {\setlength{\parsep}{0pt}%
      \setlength{\topsep}{0pt}%
      \setlength{\itemsep}{2pt}}}%
  {\end{list}}
\newcommand\Noted[1]{} 

\newcommand\xzlNote[1]{\sethlcolor{yellow} \hl{#1}} 

\newcommand\wrxNote[1]{\sethlcolor{orange} \hl{#1}} 

\definecolor{darkblue}{rgb}{0.0, 0.0, 0.55}
\definecolor{mygreen}{HTML}{ADFF2F}
\definecolor{mylightgray}{gray}{0.8}

\newenvironment{myitemize}%
  {\begin{itemize}
	[leftmargin=0cm,
		itemindent=.3cm,
		labelwidth=\itemindent,
		labelsep=0pt,
		parsep=1pt,
		topsep=1pt,
		itemsep=1pt,
		align=left]
  }%
  {\end{itemize}}    

  {\begin{enumerate}
	[leftmargin=.cm,itemindent=.5cm,labelwidth=\itemindent,
		labelsep=0pt,
		parsep=1pt,
		topsep=1pt,
		itemsep=3pt,
		align=left]
  }%
  {\end{enumerate}}    

\newcommand\sect[1]{Section~\ref{sec:#1}}	

\newcommand{\code}[1]{\texttt{\small{#1}}}	

\newcommand{\sys}{\code{PASU}}

\newcommand{\nvnano}{Jetson Orin Nano}
\newcommand{\nvorin}{Jetson AGX Xavier}

\newcommand{\allcloud}{\textit{AllOffload}}
\newcommand{\alllocal}{\textit{OnDevice}}
\newcommand{\hybrid}{\textit{NaiveHybrid}}

\newcommand{\oursno}{\textit{Ours-OnDevice}}
\newcommand{\ourslow}{\textit{Ours-Offload-L}}
\newcommand{\oursmed}{\textit{Ours-Offload-M}}
\newcommand{\ourshigh}{\textit{Ours-Offload-H}}

\makeatletter
\def\@copyrightspace{\relax}
\makeatother

\begin{abstract}

Modern speech understanding (SU) runs a sophisticated pipeline: 
ingesting streaming voice input, 
the pipeline executes encoder-decoder based deep neural networks repeatedly; 
by doing so, 
the pipeline generates tentative outputs (called hypotheses), 
and periodically scores the hypotheses. 

This paper sets to accelerate SU on resource-constrained edge devices.
It takes a hybrid approach:
to speed up on-device execution; to offload inputs that are beyond the device's capacity. 
While the approach is well-known, 
we address SU's unique challenges with novel techniques: 
(1) \textit{late contextualization}, 
which executes a model's attentive encoder in parallel to the input ingestion; 
(2) \textit{pilot inference}, which mitigates the SU pipeline's temporal load imbalance; 
(3) \textit{autoregression offramps}, 
which evaluate offloading decisions based on pilot inferences and hypotheses. 

Our techniques are compatible with existing speech models, pipelines, and frameworks; 
they can be applied independently or in combination. 
Our prototype, called \sys{}, is tested on Arm platforms with 6 -- 8 cores: 
it delivers SOTA accuracy; 
it reduces the end-to-end latency by 2x and reduces the offloading needs by 2x.

\end{abstract}


\copyrightyear{2024}
\acmYear{2024}
\setcopyright{rightsretained}
\acmConference[ACM MobiCom '24]{The 30th Annual International Conference on Mobile Computing and Networking}{November 18--22, 2024}{Washington D.C., DC, USA}
\acmBooktitle{The 30th Annual International Conference on Mobile Computing and Networking (ACM MobiCom '24), November 18--22, 2024, Washington D.C., DC, USA}
\acmDOI{10.1145/3636534.3690694}
\acmISBN{979-8-4007-0489-5/24/11}
\begin{document}

\title{Turbocharge Speech Understanding with Pilot Inference}




\author{Rongxiang Wang}
\affiliation{%
	\institution{University of Virginia}
	\city{}
    \state{}
	\country{}
}
\email{waq9hw@virginia.edu}
\author{Felix Xiaozhu Lin}
\affiliation{%
    \institution{University of Virginia}
	\city{}
	\state{}
	\country{}
}
\email{felixlin@virginia.edu}



\date{}

\thispagestyle{empty}
\maketitle

\section{Introduction}
\label{sec:intro}

Speech is a pervasive user interface for embedded devices. 
At its core are two speech understanding (SU) tasks: 
automatic speech recognition (ASR), transcribing voice to a sentence~\cite{yu2016automatic};
spoken language understanding (SLU), mapping voice to a structured intent, such as 
\code{\{scenario: Calendar, action: Create\_entry\}}~\cite{de2007spoken}.

\begin{figure}[t!]
	\centering
	\includegraphics[width=0.48\textwidth]{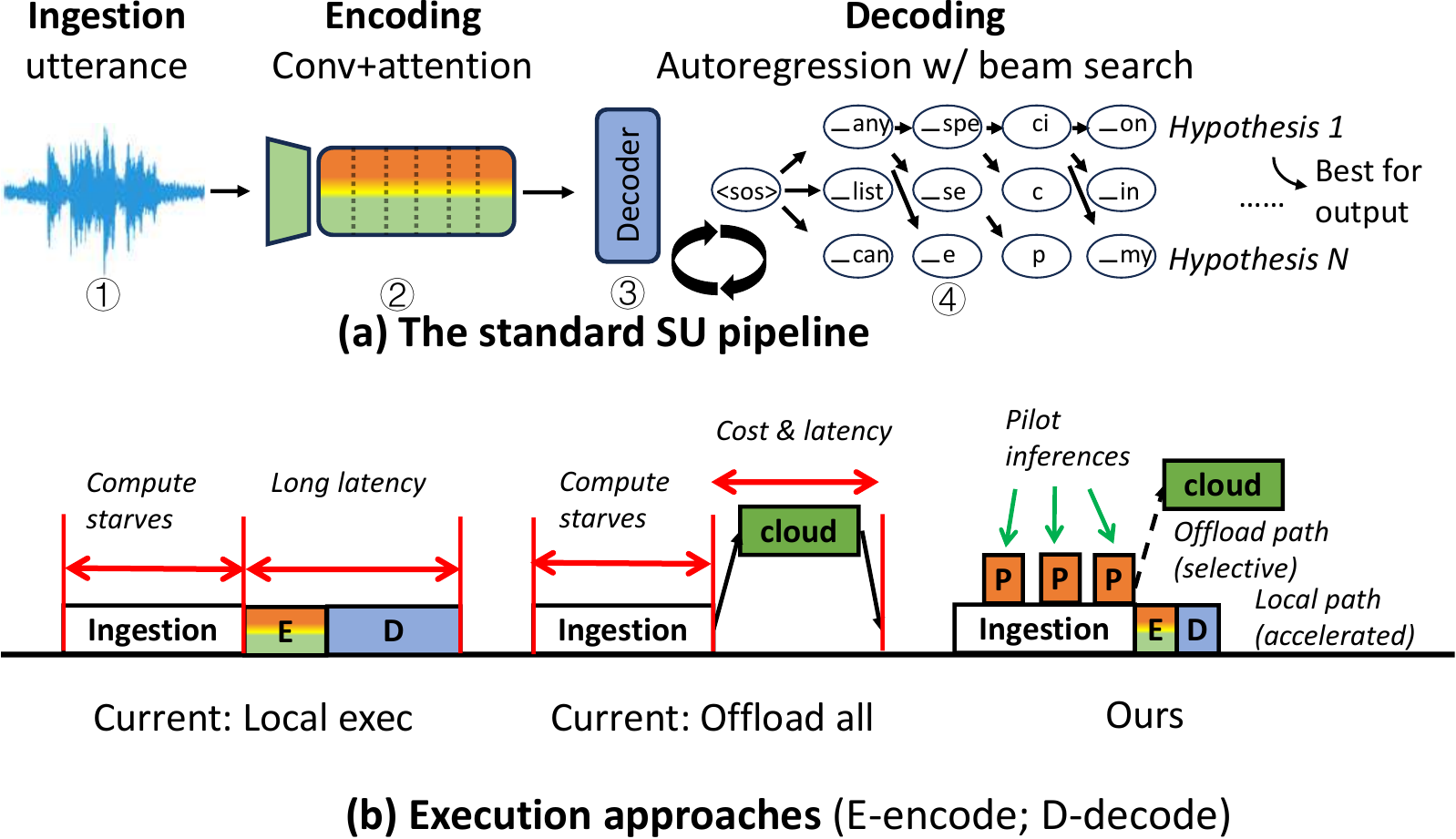}
	\caption{The SU pipeline and a comparison of execution approaches}
	\label{fig:intro}
	
\end{figure}
\begin{table}[t]
        \includegraphics[width=0.48\textwidth{}]{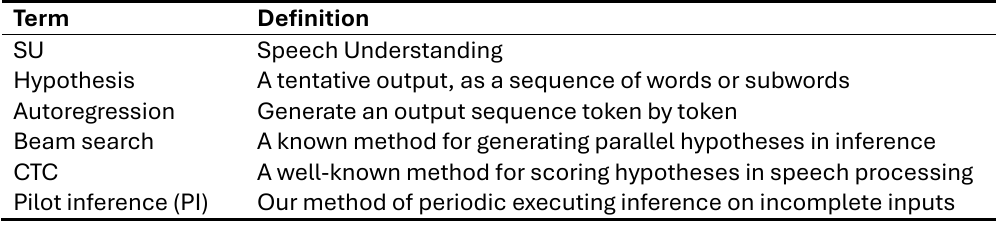}
        \caption{A glossary of terms used in the paper}
        \centering
        \label{tab:glossary}
\end{table}

\paragraph{Modern SU: autoregression with beam search}
Modern SU runs a deep, encoder-decoder model \cite{Jan2015NIPS} as shown in \autoref{fig:intro}(a). 
Ingesting an utterance waveform (\circled{1}), the model comprises a neural encoder (\circled{2}) and a neural decoder (\circled{3}), 
generating a sequence of tokens, i.e. words or sub-words (\circled{4}). 
This generation process is \textit{autoregressive}: 
to produce a token, the decoder takes as the input all the latent units from the encoder, 
as well the output tokens it produces so far. 
To cope with noisy utterances, SU runs multiple decoding processes in parallel, 
each generating a candidate output sequence (called a hypothesis). 
SU picks the best sequence as the final output. 
Thanks to the encoder/decoder wrapped in beam search, 
modern SU can understand natural utterances (e.g. ``I need to turn on light in bedroom '')~\cite{slurp_dataset} beyond simple ones (e.g. ``light on'').
This paper focuses on encoder-decoder models, referring to them as deep SU. 

Deep SU models are resource-hungry,  
often requiring GBs of memory and tens of GFLOPs per input second~\cite{arora2022two}. 
As a result, a model can take a few seconds to process a typical voice command. 
Facing the constraints, many embedded devices chose to 
offload all voices to the cloud~\cite{microsoft_azure_speech}, 
which, however, raises concerns on high cloud cost~\cite{qualcomm2023hybridcost, qualcomm2023futurehybridai, microsoft2023azurehybridbenefit}, privacy~\cite{saade2019spoken}, latency~\cite{maheshwari2018latency}, and service availability.

\paragraph{Problem \& approach}
This paper presents an engine to accelerate deep SU on resource-constrained embedded devices\footnote{shortened as ``devices'', to be defined in \sect{motiv}}.
The engine comprises a local execution path and an offloading path. 
While the local/offloading approach is reminiscent of many ML systems~\cite{shakarami2020survey,yao2020deep,lv2022walle}, our design specifically addresses unique challenges from speech. 

\subsection{The local execution path}

\paragraph{Challenge: temporal load imbalance}
We identify a key inefficiency in the Speech Understanding (SU) task: poor resource utilization during the ingestion phase. The state-of-the-art attention-based SU system necessitates the compute to remain idle during the ingestion period, creating a surge of encoding-decoding workload after receiving the complete data, as illustrated in \autoref{fig:intro}(b). Incomplete data can significantly compromise accuracy for these systems. Similar accuracy degradation is also observed in streaming SU systems, as discussed in \sect{related}.


In summary, the device needs to utilize ingestion-time compute resources, in order to accelerate the expensive processing that can only start after the ingestion.

\paragraph{Our idea: pilot inference}
During ingestion, SU periodically encodes and decodes the incomplete input the device has received so far. This is shown in \autoref{fig:intro} (b).
Our key insight is that the the pilot inference's \textit{intermediate state} can assist the full inference (i.e. the inference executed after ingestion, over the complete input). 
With the idea, we present multiple techniques: 
(1) beam collapse: 
the full inference only needs to verify its beam search path against the path in the pilot inference, and only falls back to full beam search in case of path divergence;
(2) early termination: 
extrapolating the output length of pilot inference, the full inference predicts the final output length and help with beam search early termination;
(3) CTC leap: the full inference approximates the connectionist
temporal classification (CTC) prefix scoring~\cite{watanane2017hybridctc} with the scores pre-computed by pilot inferences.

Pilot inferences are incremental in nature: 
the $(i+1)$-th inference instance reuses of the state from the $i$-th instance, 
in the same way as the full inference reuses the last pilot inference instance. 
In this way, the total cost of pilot inference is amortized over the duration of ingestion, and scales gracefully with the input length. 

Pilot inference (PI) is related to recent speculative execution for language models~\cite{kim2023speculative} with a key difference: 
while the later optimizes inference on \textit{complete} inputs, 
PI processes successive \textit{incomplete} inputs, 
for which it addresses different challenges. 
\sect{related} presents a detailed comparison. 

\subsection{The offloading path}

\paragraph{Challenge: Lacking offramps for autoregression}
Typically, 
to decide whether to offload an input $x$, 
the device assesses its \textit{confidence} on $x$, i.e. 
how likely the on-device inference yields a sufficiently accurate answer.  
Such an assessment is referred to as \textit{offramps} in prior work.
Existing offramp designs typically evaluates confidence as an ML model's prediction entropy~\cite{NEURIPS2020_d4dd111a,xin-etal-2020-deebert}. 
For SU however, existing offramps mismatch in two ways. 

First is the \textit{confidence measure}. 
Unlike a classification task which evaluate a model's layers sequentially and provides a single output that represent the probability,  
SU's output generation is both concurrent and iterative:  
it produces multiple hypotheses of probabilistic tokens sequence in a token-by-token fashion; in each iteration the decoder will generate the next tokens' probabilities base on current hypotheses.
It was unclear how to derive a single confidence score out of an ongoing generation process. 

Second is \textit{timing}. 
Given an input, offloading represents a one-shot decision that can be taken at various times throughout the SU pipeline shown in \autoref{fig:intro}. 
While a deferred decision would have more input information for measuring the model confidence, 
it also incurs on-device delays. 
The timing therefore hinges on tradeoffs between an offramp's selectivity and overhead. 

\paragraph{Our idea: PI-aid perplexity-based offloading}
The device evaluates its offloading decision as soon as it finishes ingestion, 
based on the perplexity score \cite{Jelinek1977PerplexityaMO} of the last pilot inference. 
Doing so strikes a sweet-spot: 
(1) as the discrepancy between the last pilot inference and the full inference is small, 
their perplexity scores are also similar; 
(2) the device does not need to hold off the offloading decision until the completion of full inference. 
Notably, offloading an \textit{incomplete} input (i.e. before the ingestion completes) 
does not speed up offloading the complete input: 
as an input utterance is no more than tens of KB, 
offloading is typically bound by a network round trip. With our PI-aid perplexity-based offloading, the system can achieve the target accuracy with reduced latency and offloading cost.



\subsection{Results and contributions}

We implement a system called \sys{} (Pilot-Aid SU), 
atop two embedded platforms with 6 and 8 Arm64 cores respectively. 
We test \sys{} on SLURP~\cite{slurp_dataset}, a challenging speech benchmark comprising 102 hours of speech. 
On both ASR and SLU tasks, \sys{} delivers strong accuracies with end-to-end latencies as low as 100s of ms, which is close to or below the threshold for human latency perception.
Compared to popular, efficiency-optimized on-device models \cite{conformer-2020,peng2022branchformer}, 
\sys{}'s on-device execution incurs 2x lower latency with little energy overhead. 

Combining local execution with offloading, \sys{}'s hybrid execution offers much higher accuracy than on-device models at similar or lower latency; 
the hybrid execution offers close-to-gold accuracy (only 1\% below SOTA), while reducing the offloading needs by 47\% -- 53\%.

Towards deep SU on embedded devices, we contribute:

\begin{myitemize}

\item \sys{}, a first-of-its-kind hybrid SU engine specifically optimizes for embedded devices. 

\item Pilot inference, a novel design to exploit under-utilized resources during ingestion. 

\item A new offramp design that assesses model confidence in generative, autoregressive output. 
This is the first mechanism for offloading generative tasks, to the best of our knowledge. 
\end{myitemize}
	
We will make our model and code publicly available.	


\section{Motivations}
\label{sec:motiv}

\subsection{Speech understanding (SU) at the edge}

\paragraph{An unsolved mission}
On simple utterances, 
classic ML already attained word error rate (WER) as low as 4\% on LibriSpeech~\cite{Panayotov2015librispeech}. 
On real-life speech as in the SLURP dataset~\cite{slurp_dataset}, however, WER can be as high as 20\%~\cite{Hannun2021TheHO}. 
Deep models reduce WER significantly, 
albeit requiring much more compute resources. 

\paragraph{Our system model}
A device has several CPU cores (no more than 8) and around 100 MB of memory budget~\cite{lebeck2020memoryos, soyoon2021androidmemory}.
The device is wirelessly connected to the cloud speech services; 
yet, fewer cloud invocations are preferred, 
as they incur monetary cost and increase privacy risks. 
Users expect low latency (below a few hundred milliseconds) 
after they finish speaking to the devices. 
The device detects the beginning and end of an utterance with well-known, lightweight algorithms~\cite{Liu2015AccurateEW, martin2001speechdetection}.

\begin{figure}[t!]
	\centering
	\includegraphics[width=0.45\textwidth]{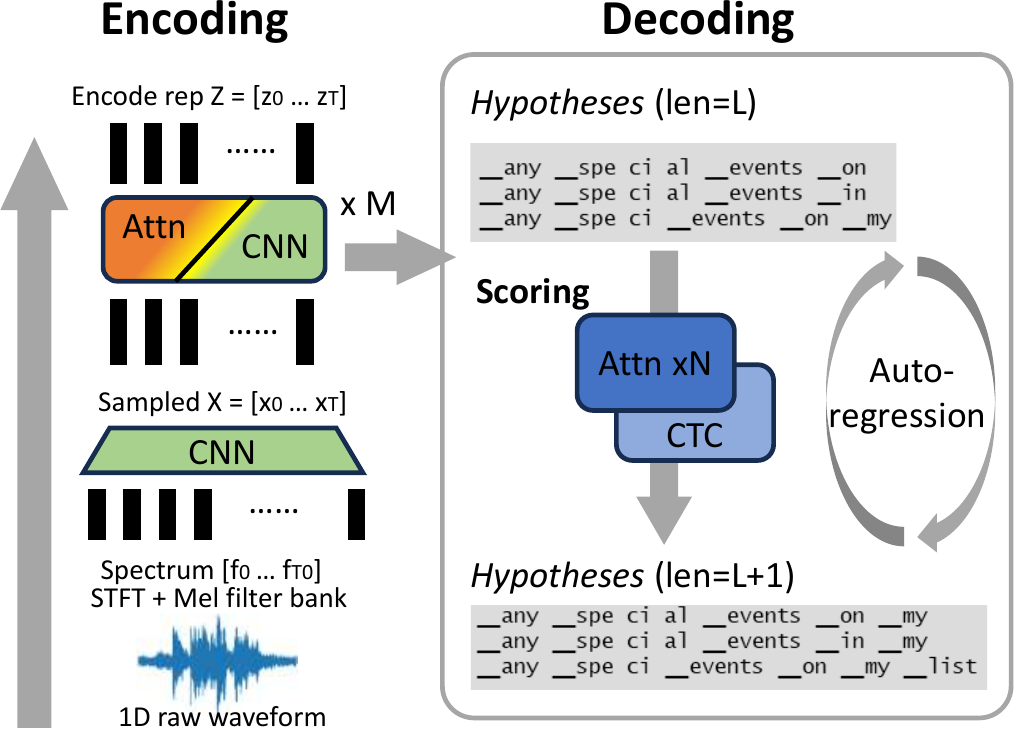}
	\caption{The deep SU pipeline~\cite{conformer-2020,peng2022branchformer,baevski2020wav2vec}.
	The input of SU tasks, an 1D raw wavform of the audio, will first go through STFT+Mel filter bank to get spectrum, then processed by encoder to get intermediate encode representation, and finally decoded in an autoregressive fashion by hybrid transformer/CTC decoder coupled with beam search method
	}
	\label{fig:motivpl}
	
\end{figure}

\subsection{A primer on SU pipelines}

As shown in \autoref{fig:motivpl}, a pipeline consists of two modules. 



\paragraph{Module 1: The encoding process}
Typically, the encoder in an SU model consists of multiple layers that combine convolution and attention-based elements. This design can be found in models such as Conformer~\cite{conformer-2020}, Branchformer~\cite{peng2022branchformer}, Wav2vec~\cite{baevski2020wav2vec}, and HuBERT~\cite{hsu2021hubert}. During operation, as shown in \autoref{fig:motivpl} left part,  the 1-D speech data will be broke in to frames and undergoes STFT and Mel-filter bank methods to create a 2D spectrum that have $f_{1} ... f_{T_o}$, each $f_{i}$ is the processed frequency component of that time frame. The 2D spectrum $[f_1, ..., f_{T_o}]$ is then processed by a few convolutional layers for downsampling, $X = x_1, ..., x_T$ is get from $X = subsampling([f_1, ..., f_{T_o}])$. The X is subsequently passes through multiple encoder layers that include both convolutional and transformer component and get the latent speech representation $Z$ among the same time frames $z_1, ..., z_T$ like X, $Z = encoders(X)$. This entire encoding process is a one-pass operation, providing an intermediate representation of the speech signal framewise with a specific dimension.

\paragraph{Module 2: The decoding process}
The generation of the final transcript involves an autoregressive beam search process as shown in \autoref{fig:motivpl} right part, with tokens generated one by one. The beam search employs a specific beam width (k), maintaining k developing hypotheses during the decoding process. In decoding, the CTC decoder first calculates posterior token probabilities $p(z_i = c | Z)$ for each frame $z_i$ of the encoder output. Beginning with the "<sos>" (start of the sentence) token, the transformer decoder calculates the next token $c$'s posterior probabilities based on current partial hypothesis $g$ and speech latent representation and get $p_{\rm attn}(c|g,Z)$, combine with the previous token probability we have $p_{\rm attn}((g,c)|Z)$. 

Throughout decoding, 
hypotheses are scored by a combination of the transformer and CTC prefix scorers~\cite{watanane2017hybridctc}. 
Prior work deem both scorers vital: the transformer exploits language features; CTC exploits acoustic features. 

Notably, the CTC scorer can be slow, as its dynamic programming algorithm shows lower parallelism, barely benefiting from modern CPU/GPU.
It calculates CTC prefix probabilities $p_{\rm ctc}((g, c, ...) | Z)$ based on newly developed hypotheses $(g, c)$ and the CTC framewise output $p(z_i = c | Z)$. The best K hypotheses with highest $\lambda \log p_{\rm attn}((g,c)|Z) + (1-\lambda) \log p_{\rm ctc}((g, c, ...) | Z)$ are retained in the beam for the subsequent rounds of decoding, based on a weighted sum of probabilities from the transformer decoder and CTC prefix scorer. The decoding process concludes when the completed hypothesis significantly outperforms the developing hypotheses, as described in~\cite{watanane2017hybridctc}.

We next analyze the inefficiencies in the two modules.

\subsection{The encoding inefficiency}


First, the ingestion is streaming. 
A typical spoken input lasts as long as 3--5 seconds ~\cite{slurp_dataset,Panayotov2015librispeech}. 
In contrast to vision and NLP systems which usually ingest complete images or sentences, 
an SU system receives an input utterance by pieces. 

\begin{figure*}[t]
 
    \includegraphics[width=\linewidth]{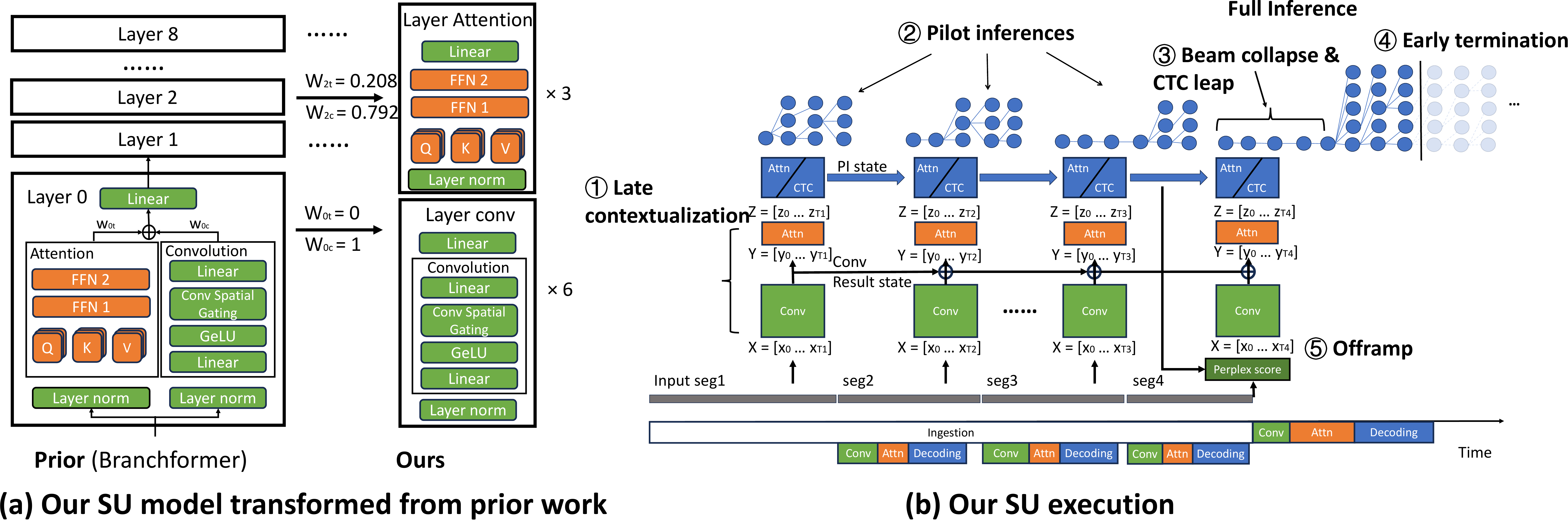}
	\caption{Overview. (a) Compared to a popular model (left), our model (right) shifts much of the encoding computation to bottom layers, allowing them to be executed in a streaming fashion during ingestion. (b) Execution pipeline of \sys{}. Our system performs pilot inference periodically on partial input, and use the pilot inference result to speedup the final inference}
	\label{fig:pl}
	
\end{figure*}

Second, the input completeness matters to accuracy. 
The information in the input audio is distributed throughout its entire duration, with the entirety of the audio serving as the context for each individual part.
While waiting for data ingestion may seem to hinder resource efficiency, it is a necessity for attention models. 
The all-to-all attention mechanism in transformer-based models is the key to their superiority in NLP and speech-related tasks. This mechanism leverages contextual information from the complete dataset to generate results with better language consistency. 
As shown in prior work \cite{Jan2015NIPS,peng2022branchformer,conformer-2020}, 
any portion of the output should take into account the whole input (i.e. the context). 
As an anecdote, in time-related phrases, words like ``calendar'' or ``reminder'' should be assigned with higher probabilities.
Partial input data with incomplete context can significantly impair the model's decoding accuracy. 
As such, a common practice is for the models to wait for ingestion completion~\cite{pundak2018deepcontext}.

As a result, throughout a deep SU pipeline the resource utilization is severely imbalanced.
The compute remains idle during ingestion, while becoming busy during the time-consuming encoding and decoding processes. 
This motivates us to overcome the aforementioned algorithmic constraint, shifting a fraction of computation to the ingestion. 
We expect only marginal increase in energy consumption, as the device is already busy processing audio IO. We will validate energy in \sect{eval}. 


\subsection{The decoding inefficiency}

Algorithmically,
SU decoding exhibits high complexity. 
If we use $NFE$ to denote 
the number of neural function evaluations in generating one token, 
the total decoding complexity is roughly $O(k\times \ell \times NFE)$, 
where the beam width $k$ is the number of parallel hypotheses, 
and the depth $\ell$ is the max hypothesis length before the search termination. 
As an example, an input of 3 seconds would require 50---125 NFEs.

Much computation is likely redundant.
Notably, the decoding process generates hypotheses much longer than the best hypothesis it eventually selected as the output: 
on popular benchmarks, we observe the average $\ell$ is \textasciitilde 18, while the true output length is \textasciitilde 11 on average, 
suggesting up to 40\% NFEs could have been saved. 




Empirically, 
the decoding is slow on typical edge devices. 
As \autoref{sec:eval} will show, a modern Arm SoC takes around 1 second to process a typical voice command, far exceeding user's perception threshold at a few hundred milliseconds. 
Unfortunately, mobile GPUs would not help much. 
Our experiments on Nvidia's Ampere mobile GPUs are even 1.6x slower than CPUs on the same board (see~\autoref{tab:platformspecs}, 6CB). 
This is likely due to the irregular decoding workloads, especially the CTC scorer, leaving the GPU hardware parallelism underutilized. 
Long decoding latency also keeps the device SoC in a high power mode, consuming more energy.

\section{\sys{} Overview}
\label{sec:overview}



\subsection{The on-device SU model}

As prerequisites to system design, 
we refactor a typical SU model. 
This is shown in \autoref{fig:pl} (a). 

\paragraph{Encoder: late contextualization}
We ensure that most of encoding layers can execute in parallel to ingestion. Rather than computing all-to-all attention at each encoding layer, we make the bottom layers (closer to the input) compute only convolution, and the few top layers compute attention.

\paragraph{Decoder}
\sys{} has separate decoder instances for pilot inferences and for the full inference. 
The two instances share the same structure and weights, whereas the pilot decoder runs a narrower search beam. 

The model details can be found in \sect{impl}.

\subsection{The system}





\paragraph{Ingestion}
As shown in \autoref{fig:pl}(b).,
\sys{} ingests segments of voice input. 
Upon the arrival of each segment, \sys{} executes the convolutional encoder layers, 
producing per-segment results (\circled{1}).
Periodically, \sys{} executes pilot inference (\circled{2}): 
combining the per-segment results accumulated so far 
and sending them to the attentive encoding layers, 
followed by the decoding process that generates a sequence of tentative tokens, 
which we refer to as a ``pilot output''. 

\paragraph{Pilot inference}
Over the course of ingesting an input, 
\sys{} repeats pilot inference multiple times, 
each time on an increasingly longer input. 
The frequency of pilot inferences is a configuration parameter to be evaluated experimentally. 
Essentially, pilot inferences extract valuable information from the partial inputs, 
\textit{during} the ingestion. 
This information speeds up full inference, which benefits the local execution path; 
it also contributes to confidence estimation, which benefits the offloading path. 

\paragraph{Offramp}
Once the ingestion is over (e.g. a detected pause~\cite{Liu2015AccurateEW}), 
\sys{} evaluates the offloading decision immediately, 
by computing the confidence score of the last pilot output (\circled{5}). 
By doing so, \sys{} makes the decision without waiting for inference on the complete input
$\{x_{{i}=1..T}\}$; 
since the last pilot inference is based on the longest partial input, 
the resultant confidence is expected to approximate to that in decoding the full input.

If the confidence is low, \sys{} offloads the input as waveform and waits for results; 
otherwise, it encodes the complete input and does full decoding locally. 
Note that: offloading intermediate results (as opposed to waveforms) often saves no network traffic, as each waveform is often tens of KBs. 
\sys{} refrains from emitting the pilot output to the user, because of high word errors.
Nevertheless, 
the pilot output benefits decoding in crucial ways: 
to reduce the beam search width (\circled{3}), to approximate the CTC prefix scores (\circled{3}), and to help with the beam search early termination (\circled{4}). 

\subsection{Applicability}

Our techniques can be applied independently or in combination to existing SU. 
(1) Late contextualization and pilot inference both speed up local execution; 
they can be applied separately to existing models. 
Specifically, pilot inference can be applied to \textit{unmodified} models, such as Branchformer and HuBERT.
Notably, it is compatible with both attention decoding (such as in Whisper~\cite{pmlr-v202-radford23a}) and hybrid CTC/attention decoding,
as we will show in \sect{eval}. 
(2) Our offramps, which introduce offloading to SU, 
can be applied to other on-device SU models in order to allow local/cloud hybrid execution. 
Without our design in (1) however, 
offloading decision would have to be evaluated on full decoding sequences, 
incurring extra delays. 

Our techniques require light modifications to the existing SU implementations. 
It does not requires to develop new ML operators. 
Late contextualization requires to train a new model architecture; the model can run atop existing ML libraries and toolkits such as ONNX runtime and ESPnet. 
Pilot inference and offramps modify the SU pipeline logic, 
which is implemented in Python code on CPU. 

Our system generally targets commodity mobile devices like smartphones and laptops. We expect these devices to be capable enough to execute the on-device model in \sys{} faster than offloading to the cloud. For more constrained devices, a customized, lighter-weight on-device model can still help orchestrate the system, though this may require compromising on local processing accuracy.




\section{The local execution path}
\label{sec:design1}

This section focuses on pilot inference for local execution.
\subsection{The mechanisms}
Pilot inference works periodically on partial data as shown in ~\autoref{fig:pl}(b) (\circled{2}). \sys{} refrains from performing pilot inference on excessively short data as it is hard to generate meaningful result with too short partial data. During data ingestion starting from $T_0$, the system conducts pilot inference on partial data periodically with a pre-defined granularity of $\delta t$, intended to complete within $\delta t$. In each round of pilot inference, the system utilizes the same encoder and the pilot decoder to engage in beam search inference with a beam size of k. Supposing the pilot inference is performed on data length $T_0+n\delta t$, it will take at most $\delta t$ to finish and the results cover the full data that have length range between $[T_0+(n+1)\delta t$, $T_0+(n+2)\delta t$]. To manage the runtime of pilot inference, a token limit of $\ell$ tokens is set. Following the decoding of $\ell$ tokens, the hypothesis with the highest probability is chosen as the reference hypothesis for local execution or hybrid execution path decisions. The reference hypothesis, denoted as $yp_{1:n} = (yp_1, \ldots, yp_n)$, offers the potential to accelerate the full inference process from multiple perspectives mentioned in the next few subsections.

\paragraph{Hyperparamters}
Several key parameters in pilot inference require pre-operation definition. These parameters include granularity $\delta t$, beam size $k$, and the maximum token length for beam search inference, denoted as $\ell$. These hyperparameters are determined through heuristic methods based on experimental experience. The granularity needs to be established initially. Typically, the system anticipates that the longest partial data covers at least ~70\% of the full-length data. To achieve this, the granularity is expected to satisfy $\delta t < 0.2T$. It's important to note that the granularity doesn't have to remain a fixed number throughout system operation. As data lengthens, the pilot inference take longer time as shown in~\autoref{fig:pl}(b). (\circled{2}) comparing the different pilot inference, the granularity can also expand. The beam size for pilot inference is set at 60\% of the beam size used for full decoding to ensure lightweight pilot inference while maintaining fidelity. The inference length is determined based on average inference length. If the average inference length is $\ell_{\rm full}$, \sys{} sets $\ell$ to be $0.7 \cdot \ell_{\rm full}$. 

\paragraph{Incremental execution}
The pilot inference based full decoding acceleration techniques (which will be elaborated in the following sub-sections) can also be adopted in the pilot inference execution. More specifically, the reference based beam collapse can be adopted in the pilot inference. In our system, as shown in~\autoref{fig:pl}(b). (\circled{2}), from the second pilot inference, each pilot inference can take reference hypothesis from the previous pilot inference and perform all the optimization in the pilot beam search procedure. Incremental execution is critical for pilot inference runtime reduction.



\subsection{Optimization 1: beam collapse} 
\textbf{Assumption} Most tokens of the full transcript should be the best token in the beam search decoding process at the corresponding round. For the beginning part of the transcript, the reference transcript from pilot inference should share most of the tokens the same with the final transcript. \textbf{What we do} Base on this assumption, during full inference, starting from the first token, in each inference round before predicting the next token probability using the scorers, the system attempts to validate the reference hypothesis. It checks the latest token in the current best hypothesis, $y_i$. If $y_i = yp_i$, the token is validated, and the system retains only the best hypothesis for the subsequent token calculation in this round. \textbf{Complexity reduction} This verification reduces the number of running hypothesis for predicting the next token to 1, also reduces required NFE to 1. \textbf{Why such approximation is acceptable} The approximation is reasonable because of two point: the reference result is from pilot inference works on most part of the data by design. The verification step also partially secures the inference not to be deviated by the potential wrong part in the reference transcript. 

\subsection{Optimization 2: early termination of beam search} 
\textbf{Assumption} The reference inference results can reflect the final transcript length, as the token number grows proportionally as time grows.
\textbf{What we do} The system predicts the final inference length using three key parameters: the length of the partial data, $\ell_{\rm partial}$ obtained in the latest pilot inference; the full data length, $\ell_{\rm full}$; and the token count of the reference hypothesis, $n_p$. The calculation for the full inference length is expressed as $n = \ell_{\rm full} / \ell_{\rm partial} \cdot n_p + C$. $C$ represents a tunable constant that loose the early termination in full inference. During the full inference, when the full inference reaches the predicted length n, if there are terminated hypotheses, the system terminates the beam search inference. Otherwise, the system continues inference until at least one hypothesis ends before terminating the process. 
\textbf{Complexity reduction} Assume the vanilla beam search length is $\ell_v$, the early termination can save $(\ell_v - n)$ round of NFE calculation.
\textbf{Why such approximation is acceptable} The approximation works well because the tokens distribute evenly among time statistically. The loose factor C can also help to decrease the deletion error that introduced by too aggressive early termination.

\begin{algorithm}[t!]
        \includegraphics[width=0.42\textwidth{}]{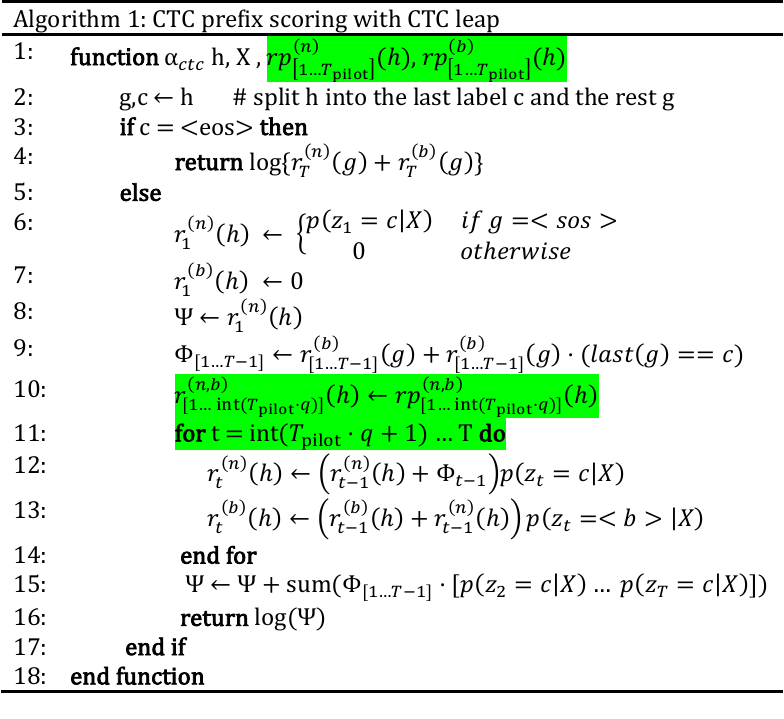}
        \caption{Algorithm modified with CTC leap for CTC prefix scoring speedup. 
        The \sethlcolor{green}\hl{highlighted} part is our modification to reuse the calculation in pilot inference and skip part of the calculation. The remaining parts are the same as the hybrid CTC/attention algorithm \cite{watanane2017hybridctc}.}
        \label{tab:ctcleap}
\end{algorithm}

\subsection{Optimization 3: fast prefix scoring with CTC leap}
\textbf{Assumption} Some of the CTC prefix scoring computation during pilot inference can be reused in the full inference process. The original CTC prefix scoring requires recursively calculated among all the time frame $[1, T]$. In pilot inference, calculation among $[1, T_{\rm pilot}]$ has been done. The system just need to finish the calculation among $[T_{\rm pilot}, T]$ 
\textbf{What we do}
We propose CTC leap to help accelerate the CTC prefix scoring step in Hybrid CTC/Attention decoding (HD) algorithm~\cite{watanane2017hybridctc} by reusing the computation in pilot inference. The algorithm modified with CTC leap is shown in ~\autoref{tab:ctcleap} The CTC prefix scoring in HD algorithm is a modified version of forward algorithm that help calculate the prefix CTC probabilities among all the time frames. In the algorithm, the prefix forward probability of prefix h among 1 to t time frames is denoted as $r_t^{(n)}(h)$ and $r_t^{(b)}(h)$, for the prefix end with non-blank and blank token. In the algorithm, the prefix h will be divided into the last token c, which is the new token that to be developed, and the prefix part g. basically (g, c) equals h. If c is the <eos> token, same with original algorithm, $r_T^{(n)}(g) + r_T^{(b)}(g)$ by definition is the final overall probability result . When c is other tokens, the $r_{[1 ... T]}^{(n)}(h)$ and $r_{[1 ... T]}^{(b)}(h)$ is calculated through the loop between line 11-14 recursively. The $\psi$ is the final cumulative prefix probability among all the time frame, calculated based on the $r_{[1 ... T]}^{(n)}(g)$ and $r_{[1 ... T]}^{(b)}(g)$ and the CTC posterior probability $p(z_t = c |X)$. In the calculation, the bottleneck is the recursive loop calculation of $r_{[1 ... T]}^{(n)}(h)$ and $r_{[1 ... T]}^{(b)}(h)$. Different from the original algorithm, as high lighted in algorithm 1, in our algorithm, we reuse the $r_{[1 ... T_{\rm pilot} \cdot q]}^{(n)}(h)$ and $r_{[1 ... T_{\rm pilot} \cdot q]}^{(b)}(h)$ from pilot inference to skip the calculation from time frame 1 to $T_{\rm pilot} \cdot q$ and only do $T_{\rm pilot} \cdot q$ to $T$ when beam collapse happens. $q\in[0.5, 1]$ is a factor that help discard the end part of the time frames. This could work because if the beam collapse happens, the reference hypothesis from pilot inference should also include prefix g, as well as the calculation result of $r_{[1 ... T_{\rm pilot} \cdot q]}^{(n)}(h)$ and $r_{[1 ... T_{\rm pilot} \cdot q]}^{(b)}(h)$.
\textbf{Complexity reduction} When the speedup requirement is met (pilot inference token is verified), the CTC prefix probability calculation in this round can be saved by $T_{\rm pilot} \cdot q / T \cdot NFE_{\rm CTC}$.  
\textbf{Why such approximation is acceptable}
We reused the $r_{[1 ... Tpilot \cdot q]}^{(n)}(h)$ and $r_{[1 ... T_{\rm pilot} \cdot q]}^{(b)}(h)$ in the pilot inference. We argue that the reused computation result is a good approximation. It is because essentially the reused calculation are (1) ultimately calculated from CTC posterior $p(z_t = c |X)$, which is calculated framewisely with CTC linear layer, thus the beginning part does not depend on the latter part. (2) the reused calculation is also based on a forward recursive algorithm, make it independent to the latter time frame.  

\subsection{Implications of approximation}
Approximate execution is common to SU and more general autoregressive system like LLM, because the whole thing is statistically in nature; and people often make approximate at their discretion, as long as the approximation can be empirically validated. The speculative decoding technique in \cite{kim2023speculative} also involves tentative hypothesis validation. The beam search end estimation technique in \cite{kasai2022beam} introduce tunable parameter "patience factor" to help tune the beam search length. The truncate CTC prefix scoring in \cite{miao2020onlinehybrid} estimate the CTC prefix alignment with time frame to save part of the computation. Different from these method, our system can better work on these estimation with the help of reference result from pilot inference.


\section{The offloading path}
\label{sec:design2}


While the above designs accelerate the local execution, 
the SU task accuracy is nevertheless bound by the model size on the device. 
To deliver SOTA accuracy, we further augment the SU pipeline with cloud execution. 
The key challenges are twofold. 
(1) The device should estimate confidence (as the offloading criteria). 
Existing method~\cite{xin-etal-2020-deebert} used in classification tasks cannot be directly applied to the generative SU task. 
It is because it works with the BERT style encoder, only take first frame of the output for classification and cannot take the full input into consideration.
(2) The device should do so without delaying the local processing. 
It cannot, for example, simply evaluate the decision \textit{after} full inference, 
as this would result in extended local decoding times for all inputs. 
It also cannot offload \textit{prior to} ingestion completion: 
as each voice input is typically tens of KB to $\sim$300 KB for speech less than 10 s, 
the offloading is bound by network round trips (RTTs); 
the device would still need one RTT to offload the whole input, after the ingestion completes.

\paragraph{Confidence estimation}
\textit{Perplexing Score-Based Approach} The perplexing score is derived from the reciprocal of the average probability of all tokens in the hypothesis, calculated by $exp(-\frac{1}{\ell}\sum_{i=0}^{\ell}\log(p(y_i|y_{0..i-1})))$. A high perplexing score indicates low confidence of the local model with the data. The system sets different thresholds for the perplexing score and offloads data that surpasses these thresholds. 

\textit{RNN-Based Approach} During beam search, the system gathers the probability of each token in the hypothesis. This fine-grained probability information better represents the local confidence in the data. Different patterns in the token probability sequence may signify different meanings. For instance, sequences of low-probability tokens often imply a lack of model confidence in the semantic correctness of an entire output span. Conversely, if such tokens are scattered, it could be attributed to acoustic noise while the output remains mostly correct. Defining rules manually can be challenging, so we adopt a learning-based approach: constructing a lightweight BiLSTM model to predict the local SU confidence. To achieve this, we label the dev set data into two classes based on their accuracy and the threshold, and then train the BiLSTM model to predict the classes. The model's output serves a similar role as the perplexing score.

\textit{CNN-Based Approach} The intermediate results from the encoder contain rich information about the data but possess significantly higher dimensions compared to the token probability sequence. To address this, we propose a CNN-based approach to assist in down sampling the intermediate results and predict the confidence. Similar to the RNN-based approach, we train the CNN model on the develop set to predict the local SU accuracy. It's noteworthy that during the design of the CNN network, we observed that the CNN model typically require a larger model size compared to the RNN approach. This issue makes it less practical. 


\paragraph{Decision timing}
The \sys{} makes the offloading decision precisely when the data ingestion is completed, along with the pilot inference result. 
This is shown in \autoref{fig:pl}(b).(\circled{5})
Both the perplexing score method and the BiLSTM method require post-decoding results. However, if the system performs confidence estimation after the full local execution, the offloaded data will also suffer from local decoding latency. To circumvent this, the system executes the aforementioned methods on the latest pilot inference result,
, i.e. the 3rd pilot inference in \autoref{fig:pl}(b).(\circled{2})
 enabling the system to make the execution path decision simultaneously with the completion of data ingestion. This approach is founded on the assumption that the partial data in the pilot inference can reflect corresponding information in the full-length data.


\section{Implementation}
\label{sec:impl}

We implemented \sys{} using 4K SLOC in Python, built upon a popular speech toolkit ESPnet \cite{watanabe2018espnet}
. 

\paragraph{Model design and training details}
Our model is motivated by the Branchformer \cite{peng2022branchformer}, as shown in the comparison in \autoref{fig:pl}(a). Branchformer has parallel weighted sum convolutional and transformer blocks in each layer for local and contextual feature extraction. 
Our model keeps the blocks that have higher weight and rearrange them to make the convolutional layers at the bottom (near the input). 
To train our model, we first configure and train an example branchformer model base on the memory and latency budget. In the experiment case, encoder has 9 layers and decoder has 1 layer. To construct \sys{}'s model,  assume there are k transformer blocks with weights exceeding $h = 0.2$ in the trained branchformer encoder,  we design the encoder with N-k-1 CNN layers followed by transformer k+1 layers. In the experiment, we get 2+1 transformer and 9-2-1 CNN layers. This design is then trained from scratch on the same dataset, with the same decoder design branchformer model has. 


\paragraph{Operation details}
During system operation on the SLURP dataset, we set the shortest partial data for pilot inference at 1.5 seconds. The granularity is determined based on hardware specifications, tested across 0.5s - 2s. The pilot beam size is set to be 3, which is 60\% of the full inference beam size of 5. The length limit for pilot inference tokens is set to be 15. 
The length prediction $C$ is set to be 5. 

\paragraph{Hybrid execution details}
For hybrid execution and offloading decision-making, in the perplexing score-based method, we calculate the perplexing score based on the reference result from the pilot inference. Regarding the LSTM and CNN-based methods, we utilize the full dev set in SLURP as the training set for the LSTM and CNN models. The CNN approach employs a CNN model that takes the encoder's full output as input. The CNN model has two convolutional layers and two fully connected layers, with 192.5 M parameters. The LSTM approach employs a BiLSTM model that takes CTC prefix probability and transformer token-wise probability of the reference result as the input. The LSTM model we adopted is a 1-layer BiLSTM model with 11.1k parameters. The training task is defined as a binary classification task, where a local execution WER > 0.1 is labeled as 1, and otherwise as 0. During operation, the BiLSTM and CNN process the respective input and provide predictions, representing the probability of the data having a WER > 0.1 as a float number between 0 and 1. Further, we set thresholds on this output for different offloading accuracy targets.

\section{Evaluation}
\label{sec:eval}
We answer the following questions: 
\begin{myitemize}
    \item Can \sys{} reduce latency with competitive accuracy?
    \item \sys{}'s local path: what is its efficacy? 
    \item \sys{}'s offloading path: what is its efficacy? 
\end{myitemize}

\subsection{Methodology}
\paragraph{Test platforms}
As shown in \autoref{tab:platformspecs}, we test on two embedded boards: 
\nvnano{} (shortened as Orin) with an efficiency-oriented SoC with six cores, whereas 
\nvorin{} (shortened as Xavier) with a performance-oriented SoC with eight cores. 
Our experiments focus on CPUs, though our idea applies to GPUs as well. 

We run the cloud runtime on an x86/NVidia machine and measure accuracy. 
To better estimate the cloud offloading delay, 
we measure Microsoft's speech service~\cite{microsoft_azure_speech}: 
we invoke the Azure APIs to offload waveforms and record each waveform's end-to-end wall time. 
Our measurement runs from the US east coast and invokes datacenters on the east coast. 
For each input, we repeat the test on enterprise WiFi (bandwidth in 100s of MB and RTT in ms) and on 5G  networks, and take the average. 
Note that we do \textit{not} use Azure speech for accuracy evaluation, as we find its accuracy inferior to the SOTA model used by us (two-pass SLU \cite{arora2022two}). 

\begin{figure*}[t!]
  \centering

    \includegraphics[width=\linewidth]{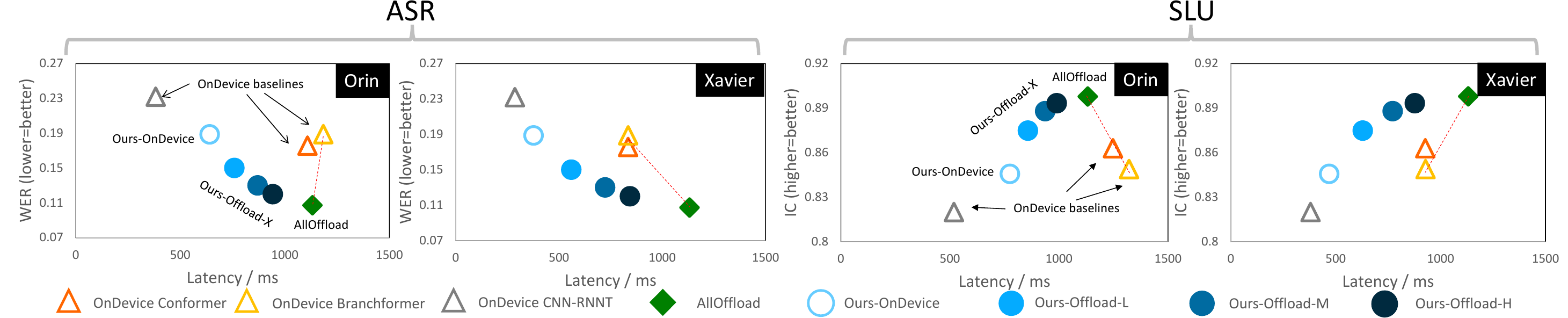}
  
	\caption{Our system deliver low delays and competitive accuracy. 
	Compared to other local models \textbf{marked with} \textcolor[rgb]{1,0.4,0}{$\bigtriangleup$}\textcolor[rgb]{1,0.75,0}{$\bigtriangleup$}\textcolor[rgb]{0.5,0.5,0.5}{$\bigtriangleup$}, our local execution Ours-OnDevice \textcolor[rgb]{0.4,0.8,1}{$\bigcirc$} show similar accuracies at much lower delays or higher accuracies; 
	Compared to \textit{AllOffload} \textcolor[rgb]{0,0.5,0}{\rotatebox[origin=c]{45}{\rule{0.5em}{0.5em}}}, our hybrid executions\raisebox{-0.6ex}{\textcolor[rgb]{0.016,0.67,1}{\scalebox{2.2}{$\bullet$}}\textcolor[rgb]{0,0.42,0.63}{\scalebox{2.2}{$\bullet$}}\textcolor[rgb]{0,0.165,0.247}{\scalebox{2.2}{$\bullet$}}}show much lower delays with little accuracy loss.}
	\label{fig:e2e}
\end{figure*}

\begin{table}[t!]
    \centering
        \includegraphics[width=0.45\textwidth{}]{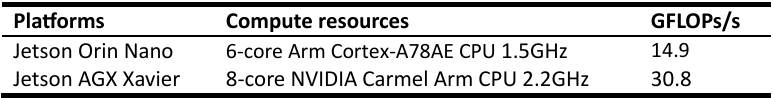}
        \caption{Embedded platforms used in evaluation}
        \label{tab:platformspecs}
\end{table}

\begin{table}[t!]
    \centering
        \includegraphics[width=0.45\textwidth{}]{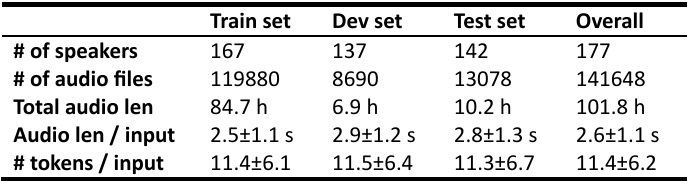}
        \caption{Overview of SLURP dataset for SU tasks}
        \label{tab:slurpdataset}
\end{table}

\paragraph{Dataset and model training}
We run our experiments on SLURP~\cite{slurp_dataset} as summarized in \autoref{tab:slurpdataset}. 
Comprising 102 hours of speech, 
SLURP's utterances are complex and closer to daily human speech
(e.g. ``within the past three months how many meetings did i have with mr richards''),
as opposed to short commands (e.g. ``light on'') in many other datasets. 
Of all 141,530 utterances, we use 119,880 (85\%) for training and the rest for dev set and test set. 
The accuracy and latency are reported from the test set data range from 2s to 6s. The on-device model training was performed on an RTX 2080 Ti GPU and took 48 hours. It is important to note that this training is a one-time effort.

\paragraph{Metrics}
We follow the common practice. 
\underline{Accuracy.}
For ASR, we report word error rate (WER): 
the discrepancy between the model output and the ground truth transcript, 
 defined as count of errors (substitutions, deletions, insertions) normalized by the sentence length.  
 For SLU, we report the intent classification accuracy (IC). 

\underline{Latency.}
(1) We report user-perceive latency~\cite{yu2016automatic}: 
the elapsed time from a user completing her utterance to the system emitting the complete SU result. 
(2) We further report the real time factor (RTF): 
the user-perceived latency normalized by the input utterance's duration.

\paragraph{Baselines}
We compare the following designs. 

\begin{myitemize}
    \item \alllocal{}: SU completely runs on device, for which we test a wide selection of 
    model architectures: 
    Conformer~\cite{conformer-2020}, Branchformer~\cite{peng2022branchformer}, and 
    CNN-RNNT~\cite{zhang2020transformer}. 
    Note that CNN-RNNT is streaming, with 640ms chunks as in prior work. 
    For fair comparison, we choose models sizes to be around 30M parameters, which are comparable to ours. 
	They are summarized in \autoref{tab:modelflops}.     
    
    \item \allcloud{}: The device offloads all inputs, 
    for which the cloud runs a SOTA model~\cite{arora2022two}: 
	a two-pass end-to-end SLU model with one conformer-based encoder, 
	one conformer-based deliberation encoder, 
	two transformer-based decoders for the two passes, and a pretrained BERT language model.
	It has >10x parameters as compared to the local models above. 
    We refer to its accuracy as \textit{gold}. 
    
    \item \hybrid{}:     
    To combine \alllocal{} and \allcloud{}, 
    execute any input with probability $\alpha$ for local execution and $(1-\alpha)$ for offloading.    
    
    
    \item \textit{Ours}: 
    By varying the confidence threshold $\theta$, 
    we tested a range of variants which we refer to as     
    \textit{Ours-OnDevice} (0\% offloaded), 
    \textit{Ours-Offload-L} (25\%), 
    \textit{Ours-Offload-M} (47\%), 
    and \textit{Ours-Offload-H} (63\%). Here the offloading thresholds and percentages are chosen to meet overall WER targets 0.15, 0.13 and 0.12.

\end{myitemize}

\subsection{End-to-end results}

As demonstrated in Figure~\ref{fig:e2e}, 
\sys{} is able to deliver a wide range of accuracies (from modest to nearly gold), 
with latencies as low as 100s of ms. 


Compared to \alllocal{}, all variants of our system are better in the latency - accuracy trade-off. 
\oursmed{} and \ourshigh{} achieve much higher accuracy (WER lower by 5\%; IC higher by 2.5\%) at similar or lower latencies; 
\oursno{} achieves similar accuracies while reducing latencies by around 2x.

Compared to \allcloud{},
\ourslow{}  and \oursmed{} reduce the overall latency by as much as 
37\% and 27\% while seeing minor degradation in accuracy 
(no more than 4\% WER and 2\% of IC). 
They process 75\% and 53\% of inputs on device, 
reducing the need for cloud invocations by 2x or more. 
\ourshigh{} has negligible accuracy degradation (<0.5\% of IC)
while still reducing latency by 20\% and offload needs by 37\%. 

\hybrid{}, 
shown as the dash lines interpolating between \alllocal{} and \allcloud{} on \autoref{fig:e2e}, 
always shows inferior accuracy or latency to ours. 


\subsection{Efficacy of our local path}


As shown in \autoref{fig:e2e}, 
our local-only execution (\oursno{}) significantly outperforms \alllocal{}, 
reducing latencies by 1.7x - 2.2x at similar accuracies.
We next break down the benefit into two parts: encoding and decoding. 



\begin{table}[t!]
    \centering
        \includegraphics[width=0.48\textwidth{}]{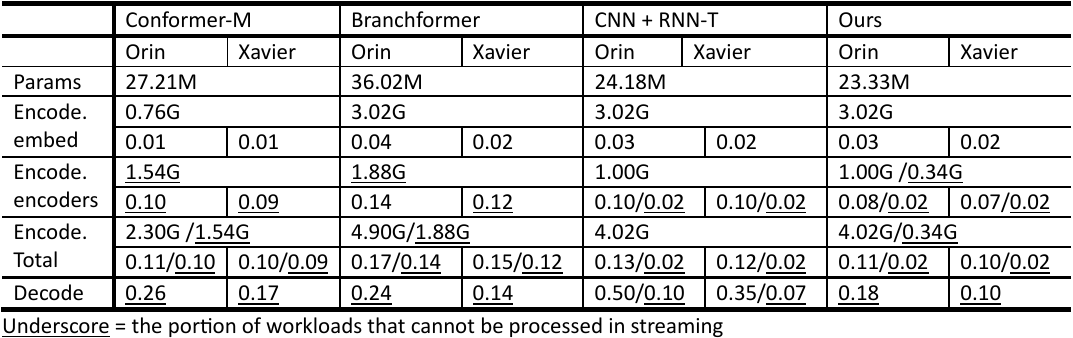}
        \caption{Our local encoder executes mostly streaming FLOPs, 
        for which latency can be hidden behind IO; 
        doing so does not compromise the final accuracy. 
        FLOPs numbers normalized to 1 second of input.
        Our local execution path incurs lower latencies (reported as RTF) as compared to other on-device models. 
        }
        \label{tab:modelflops}
\end{table}

\paragraph{Encoding speedup}
We strike a sweet spot between speed (by processing inputs in a streaming fashion) 
and accuracy (by attending to the whole input). 
\autoref{tab:modelflops} breaks down the encoding computation. 
Compared to encoders with all attention layers (Con/Branchformers)
where 67\% and 38\% FLOPs are non-streaming (i.e. must execute \textit{after} ingestion),
92\% FLOPs of ours is streaming, for which the latency is hidden behind the ingestion. 
Meanwhile, we do not lose much accuracy compared to Con/Branchformers, 
as our design runs three attention layers after ingestion ends. 
This design reduces the latency by 3.6x - 5.4x (220 -- 345 ms or 0.07 -- 0.11 RTF) as shown in \autoref{tab:modelflops}.
Compared to full streaming encoders (CNN+RNNT) for which 100\% FLOPs is streaming, 
our accuracy is much higher by 4\% in WER and 2.6\% in IC, 
because the former critically lacks attention over the whole input. 
Meanwhile, our encoding latency is only higher by 0.005 RTF (15 ms for a 3 second input) as in \autoref{tab:modelflops}. 
Such a difference is dwarfed by the difference in decoding delays. 

\begin{figure}[t!]
	\centering
	\includegraphics[width=0.48\textwidth]{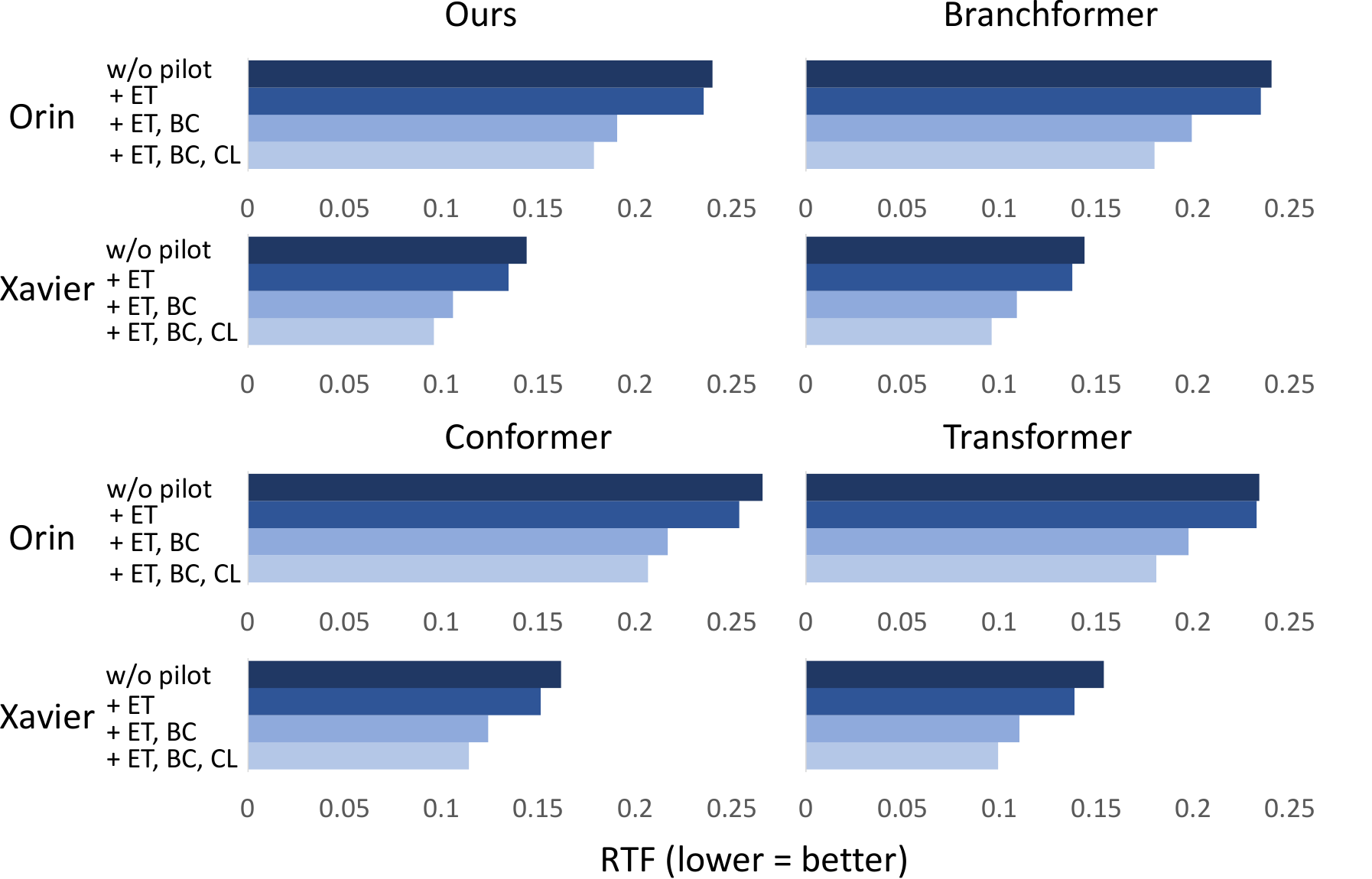}
	\caption{An ablation study of \textit{pilot inference}, showing that all its three optimizations contribute to lower latency significantly. It also shows that pilot inference has generalizability to be applied to various on-device models (e.g. Branchformer, Conformer, and Transformer) other than ours. ET: Early termination, BC: Beam collapse, CL: CTC Leap}
	\label{fig:speedupablation}
\end{figure}
\paragraph{Decoding speedup}
Our design shows 1.5x lower decoding delays compares to \alllocal{} (\autoref{tab:modelflops}).
All three techniques show contribution to the overall speedup as \autoref{fig:speedupablation} shows. We also demonstrate the decoding speedup method on branchformer. The speedup ratio is similar with it on ours model, shows that our speedup method is applicable on general transformer based encoder-decoder speech model.

The \textit{incremental}  nature of pilot inference is crucial, 
as it amortizes the decoding cost over the past input, making the cost scale slower as the input grows. 
Turning off the incremental design (i.e. each pilot inference starts from the input's start) increases the average decoding delay by 30\%, from 156 ms to 205 ms;
it increases the 90th percentile delay by 40\%, from 198 ms to 282 ms. 
Reduction in pilot inference cost allows the system to execute pilot inference more frequently. 
The pilot inference shows generalizability on mainstream SOTA speech models. As \mbox{\autoref{fig:speedupablation}} shows, besides our model, the pilot inference can effectively bring speedup to Branchformer, Conformer (with transformer decoder) and Transformer model. All three techniques contribute to the overall latency reduction.

\begin{table}[t]
    \centering
        \includegraphics[width=0.48\textwidth{}]{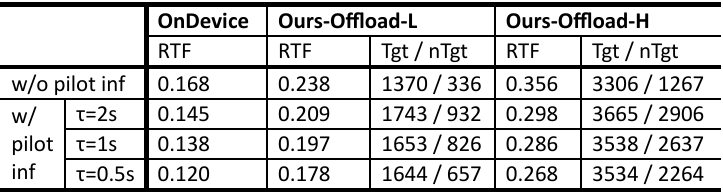}
        \caption{Pilot inference reduces the end-to-end delay. 
        It benefits our local-only execution (Ours-NoOffload) as well as our hybrid executions (Ours-Offload-X),
        making the latter's offloading more selective. Tgt / nTgt refers to target / non-target inputs, respectively, in which Tgt are the input data on which the cloud gets lower WER, i.e. they should be offloaded.}
        \label{tab:psu}
\end{table}

We further study the impact of pilot inference's eagerness:
during ingestion, every $\tau$ seconds \sys{} decodes the input it has accumulated so far. 
Lower $\tau$ reduces the discrepancy between the last pilot inference and the full decoding, 
therefore improving the full decoding speed and quality.
As shown in \autoref{tab:psu}, 
4x reduction in $\tau$ (from 2s to 0.5 s) reduces final RTF by 0.025.
further reduction in $\tau$ helps the pilot inference get longer partial data that cover more part of the full length data
In exchange, the expense is that the ingestion consumes more compute. 
$\tau$  is lower bounded by the available compute resource during ingestion. 
For instance, \nvorin{} can sustain $\tau=0.5s$ while \nvnano{}  cannot. 

\begin{table}[t]
    \centering
        \includegraphics[width=0.48\textwidth{}]{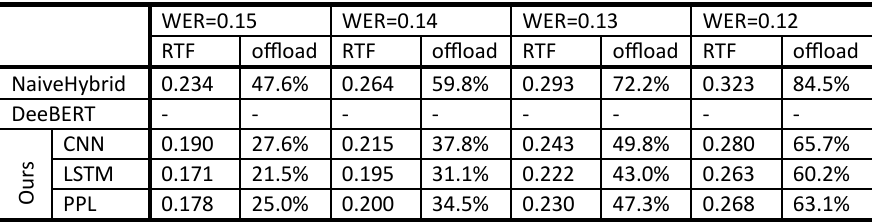}
        \caption{Our offloading strategies based on sequence modeling outperform  \textit{NaiveHybrid} (random selection) and DeeBERT (sequence classification, failing to produce useful results). 
        In the experiment, we tune $\theta$ and $\alpha$ to meet WER goals and compare delays and offloading ratio}
        \label{tab:offloadingstrategy}
\end{table}

\begin{figure}[t!]
	\centering
	\includegraphics[width=0.48\textwidth]{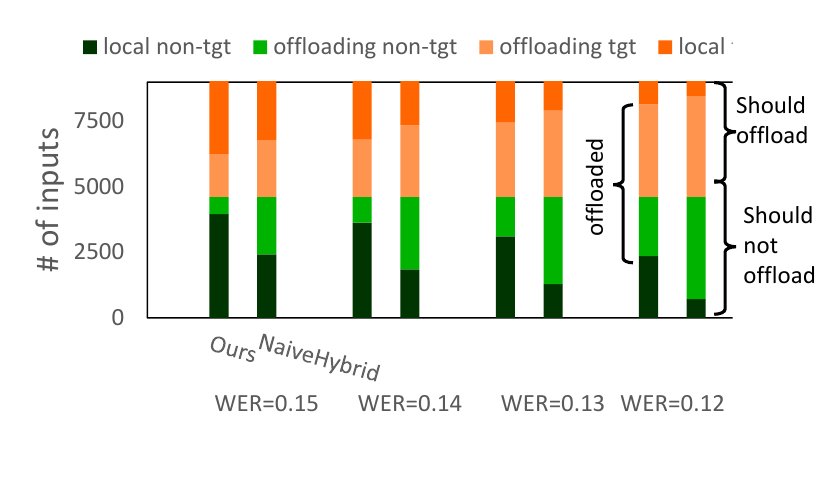}
	\caption{Our design shows good selectivity in making offloading decisions, 
	offloading much fewer inputs compared to \textit{NaiveHybrid} achieving similar accuracies.	tgt / non-target refer to target / non-target input, target input is data that get lower WER on cloud
	}
	\label{fig:obd}
	
\end{figure}
\subsection{Efficacy of our offloading path}

\sys{} effectively identifies and uploads the inputs 
that would suffer from low accuracy on device. With our selective offloading technique, \mbox{\sys{}} achieves lower latency and offloading cost. In this section, we define data as target inputs (shortened as Tgt) when they get lower WER (i.e. higher accuracy) from the cloud processing than local processing.
 
 
\paragraph{Comparison vs. \textit{NaiveHybrid}}
We replace \sys{}'s offloading strategy with \textit{NaiveHybrid} 
while keeping all other optimizations. 
The results in \autoref{tab:offloadingstrategy} show that
to reach the same accuracy (WER), 
\textit{NaiveHybrid} has to offload up to 2x more inputs. 
The extra offloading translates to higher cloud cost 
as well as higher latency (0.06 RTF; 180 ms on average). 

\autoref{fig:obd} shows detailed offloading decisions: 
For \oursmed{} (WER=0.13) and \ourshigh{} (WER=0.12), 
\sys{} offloads the majority of target inputs, 
while still executing the majority of non-target inputs on device. This is much higher than \textit{NaiveHybrid} which make decisions ``by chance''.


\paragraph{Comparison vs. DeeBERT}
Our experiment also shows that 
a popular early-exit approach~\cite{xin-etal-2020-deebert}, 
estimating model confidence with linear classifiers inserted after the encoder,  
performs poorly for encoder-decoder SU models. 
We implemented such an approach: 
training linear classifiers atop the first output frame embedding from the encoder. 
We could not get meaningful results 
-- the classifier is no better than a random predictor. 
This highlights the challenge of predicting SU confidence,
for which the entire generated sequence (not just the 1st frame from encoding) must be considered. 


\paragraph{Choice of sequence modeling strategies}
The sequence modeling techniques in \sect{impl} (CNN, LSTM, and perplexity) can well estimate the model confidence. 
\autoref{tab:offloadingstrategy} shows that LSTM offers the lowest offloading percentage and RTF. PPL perform slightly worse, but 
as perplexity has much lower computational complexity (no training, no parameters), 
we deem PPL as the most suitable. 

\paragraph{How PI affects offloading decisions?}
Recall that \sys{} makes offloading decisions based on the last pilot inference (\autoref{sec:design2}). 
We answer the following questions.
(1) How does pilot inference's execution plan affect offloading decisions? 
Our results show that 
finer granularity (i.e. smaller discrepancy between the pilot and the full inference) 
leads to better estimation of model confidence, which results in more accurate offloading decisions. 
For instance in \autoref{tab:psu}, to maintain the accuracy at WER=0.15, 
reducing $\tau$ from 2 s to 0.5 s reduces
the offloaded inputs by 14\%, which translates to 93 ms lower end-to-end delay. 
%
(2) 
What if we make offloading decisions based on the \textit{full} decoding outcome? 
Our results show that while the offloading selectivity slightly improves, the end-to-end latency is much higher (by 264 ms or 0.088 RTF), 
because the device makes offloading decisions much late -- 
after ingesting the whole input and executing full decoding. 

\subsection{Energy impact} 

\begin{figure}[t!]
	\centering
	\includegraphics[width=0.48\textwidth]{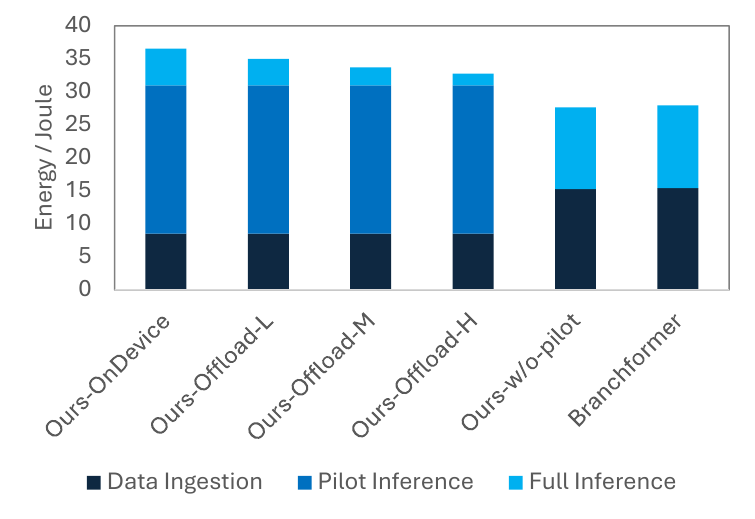}
	\caption{Our system incurs minor energy overhead (18\% to 32\%) as compared to the baseline system.}
	\label{fig:energy}
	
\end{figure}

\paragraph{Methodology}
We study system-level energy consumption to demonstrate the impact of our techniques, specifically focusing on the overall energy consumption during speech processing. The experiments are conducted on \mbox{\nvorin{}} with the built-in power profiling tool tegrastats. Power profiling is performed during inference on all test data, with a measurement interval of 20ms to capture the device's overall power consumption. Six different settings are tested: four on \mbox{\sys{}}, include \mbox{\alllocal{}}, \mbox{\ourslow{}}, \mbox{\oursmed{}}, and \mbox{\ourshigh{}}. The remaining two settings involve the default full encoding-decoding execution scheme with our on-device model and the branchformer model. The average energy consumption on each voice input is shown in \mbox{\autoref{fig:energy}}, with a detailed breakdown of energy consumption across three execution stages: data ingestion, pilot inference (if applicable), and full inference.
\paragraph{Energy comparison}
In the experiments mentioned above, we observed a {32\%} energy overhead on our \mbox{\alllocal{}} system compared to the two default full encoding-decoding systems with ours on-device model and the branchformer model. The average inference energy consumption breakdown is shown in \mbox{\autoref{fig:energy}}. From the figure, we can see that although the pilot inference introduces energy overhead, the full inference is shorter, therefore reducing the overall energy consumption overhead. With the selective offloading, \mbox{\sys{}} can further reduce the energy consumption overhead by up to {44\%}. It's important to note that the \mbox{\nvorin{}} used in the experiments is relatively power inefficient. We assume that more efficient modern chips, such as Apple M1 and A17, would result in a lower energy overhead.
\paragraph{What-if analysis}
Many factors may influence the overall energy consumption. 
Assuming we have larger models,  \mbox{\sys{}} will see greater latency reduction, which therefore saves more energy in the full inference process. Designing a more energy-efficient model can also help reduce the overhead of \mbox{\sys{}} in the pilot inference stage. Also, as mentioned earlier, more efficient hardware processors can help mitigate the energy overhead of pilot inference.




\subsection{Discussion} 
\paragraph{Device hardware}
As mentioned in \mbox{\sect{overview}}, our system targets commodity devices capable of executing the local model notably faster (e.g. >50\%) than offloading to the cloud, in order to warrant the higher local WER (often by 7\%-10\%). Less capable devices, however, experience longer latency, making it impractical compared to simply offloading to the cloud. The problems are twofold: constrained hardware leads to longer pilot inference latency and makes the intervals between consecutive pilot inferences longer, impairing its efficacy for both local and offloading paths. Additionally, the resource constraints also slow down the full decoding. Experiments on a Raspberry Pi 4b with the same model and system settings showed that the pilot inference intervals cannot be less than 1.2 seconds, otherwise some of the pilot inference could not finish before the next pilot inference starts. The overall latency was 0.343 RTF, only marginally faster than cloud offloading. While our system can further compress the model to suit the device resource constraints, this would also result in a drop in accuracy. 

Within the capable device regime, 
our local execution maintains consistent speed advantages over baselines, 
as shown by the comparison \nvnano{} and \nvorin{} in \autoref{fig:e2e}. 
Besides, faster device computation can perform pilot inference with smaller interval. 
\paragraph{Network conditions}
Our measurement was from decent network conditions that favor offloading. 
With longer network delays, 
\sys{}'s benefit would be even more pronounced, as fast local executions and selective offloading will be more important. 
If network delays further reduce, the latency gap between our system and  \textit{OffloadAll} may narrow. 
Even if that happens, our benefit of reduced cloud cost will still remain. Network conditions, especially when unstable, could also be a key consideration in offloading decisions. The system may need to adjust the offloading threshold based on network latency and availability.
Under typical network conditions, placing the SU model on a nearby edge system may not improve overall performance. This is because the SU model on the cloud is resource-intensive, running it on the edge could result in a latency increase that is significantly longer than the network roundtrip latency.





\section{Related work}
\label{sec:related}


\paragraph{LLM speculative decoding (SD)}
Speculative decoding for LLM decoding speedup includes a fast generating and verify later pattern. Examples include copy-and-verify scheme ~\cite{yang2023inference}, non-autoregressive translation methods ~\cite{xia2022lossless}, shallow decoder and parallel decoding~\cite{sun2021instantaneous}, token confidence based decoding~\cite{kim2023speculative}. Similar techniques have been adopted in speech processing, such as the Whisper Speculative Decoding \mbox{~\cite{huggingface_whisper_speculative}} in which the audio is first encoded with a transformer encoder, then decoded by a distilled, lightweight decoder initially, and later verified by a larger decoder through a one-pass forward process.
\textbf{Comparing SD to our pilot inference} 
 \textit{Tasks:} Pilot inference targets streaming SU task, addressing temporal workload imbalance. 
 SD targets LLM and SU decoding, tackling underutilized hardware parallelism. 
\textit{Solutions:} Pilot inference work with incomplete streaming input. SD deals with a complete input with concurrent fast and slow tasks. SD's speedup is contingent upon high hardware parallelism, which is lacking on embedded devices. The connection in between is that the full decoding benefit from an earlier decoding process. More specifically, Whisper Speculative Decoding still encodes the complete audio and decodes with two-pass speculative scheme all after the ingestion finished. In contrast, our system exploits the temporal load imbalance, finishs part of the encoding and pilot inference during the audio ingestion and saves a substantial amount of time.

\paragraph{Streaming speech procssing} 
Streaming approaches like 
transducer-based models~\cite{graves2012sequence,rao2017exploring,zhang2020transformer} process the data in a chunk-wise fashion, initiating transcription during data ingestion and reduces the latency. However, they only possess partial data during streaming processing and lack contextual information. Compare to attention-based model~\cite{baevski2020wav2vec,conformer-2020,peng2022branchformer}, the accuracy of streaming models is inferior ~\cite{li2020comparisone2emodel}. Our work focuses on speedup attention-based models.





\paragraph{Offloading and hybrid execution}
Device-cloud collaboration~\cite{mach2017mobileedgecomputing} is well-discussed in general scenarios. The cost during collaboration is an important concern~\cite{qualcomm2023hybridcost,qualcomm2023futurehybridai,microsoft2023azurehybridbenefit}. A related work optimizes SU for microcontrollers in the local/cloud setting~\cite{benazir2023leveraging}. Yet, incapable of deep SU models 
, microcontrollers only run an inference \textit{cache}, rather than a complete engine like \sys{}. These two projects have orthogonal contributions; they do not depend each other.  
\paragraph{Comparing CTC Leap with the Online CTC/Attention Method} 
Online CTC/attention method ~\cite{miao2020onlinehybrid} reduces CTC scoring latency by Truncated CTC method. Both Truncated CTC and CTC leap try to skip part of the CTC prefix scoring computation among all the time frame. The difference is that: we reuse the CTC computation from the pilot inference to skip the beginning part; whereas they stop early by using the CTC peak heuristics to skip the latter part. Notably, same with the original approach ~\cite{watanane2017hybridctc} our method still includes all the time frames while they choose to drop some temporal information, which may include crucial information for CTC. 

\paragraph{Mobile speech processing}
Various efforts have been made to run SU tasks on resource-constrained mobile devices. LUT-NN \mbox{~\cite{tang2023lutnn}} employed table lookup inference execution for machine learning models to enhance inference speed and reduce memory footprint. However, the lookup table scheme is specifically designed for the encoding process and cannot be easily applied to generative tasks in SU. Other methods such as quantization \mbox{~\cite{gondi2022wav2vec20edgeperformanceevaluation, xu2021Quantization}} and pruning \mbox{~\cite{lai2021parp, Peng2023structuredpruning}} have also been proposed. These methods are orthogonal to our approach. 
\section{Conclusions}
We present \sys{}, a novel system for fast speech processing in SU tasks. \sys{} contributes late contextualization, beam collapse/ termination, and CTC leap for local execution; 
and confidence estimation for selective offloading. 
All optimizations combined, \sys{} speeds up its local path by 2x and reduces the offloading needs by 2x in its offloading path.


\section*{ACKNOWLEDGMENT}
The authors were supported in part by NSF awards \#2128725, \#1919197, \#2106893, and Virginia’s Commonwealth Cyber Initiative. The authors thank the anonymous reviewers for their insightful feedback. 


\bibliographystyle{plain}
\bibliography{bib/abr-short,bib/wrx}



\end{document}